\documentclass[a4paper,fleqn]{cas-sc}

\usepackage[numbers]{natbib}
\usepackage{multirow}
\newcommand{\tabincell}[2]{\begin{tabular}{@{}#1@{}}#2\end{tabular}}
\usepackage{listings}
\usepackage{xcolor}
\usepackage{setspace}

\definecolor{codegreen}{rgb}{0,0,0}
\definecolor{codegray}{rgb}{0,0,0}
\definecolor{codepurple}{rgb}{0,0,0}
\definecolor{backcolour}{rgb}{1,1,1}

\lstdefinestyle{mystyle}{
  backgroundcolor=\color{backcolour},   commentstyle=\color{codegreen},
  keywordstyle=\color{codegreen},
  numberstyle=\tiny\color{codegray},
  stringstyle=\color{codepurple},
  basicstyle=\ttfamily\footnotesize,
  breakatwhitespace=false,         
  breaklines=true,                 
  captionpos=b,                    
  keepspaces=true,                 
  numbers=left,                    
  numbersep=5pt,                  
  showspaces=false,                
  showstringspaces=false,
  showtabs=false,                  
  tabsize=2
}

\usepackage{subfigure}

\def\tsc#1{\csdef{#1}{\textsc{\lowercase{#1}}\xspace}}
\tsc{WGM}
\tsc{QE}
\tsc{EP}
\tsc{PMS}
\tsc{BEC}
\tsc{DE}
\begin{document}
\let\WriteBookmarks\relax
\def\floatpagepagefraction{1}
\def\textpagefraction{.001}
\shorttitle{\textbf{swdpwr}: Power Calculation for SWD}
\shortauthors{J Chen et~al.}

\title [mode = title]{\textbf{swdpwr}: A SAS Macro and An R Package for Power Calculation in Stepped Wedge Cluster Randomized Trials}                      



\author[1]{Jiachen Chen}[orcid=0000-0002-1941-2326]
\ead{chenjc@bu.edu}

\address[1]{Yale University School of Public Health, New Haven, CT 06511, United States}

\author[1]{Xin Zhou}
\ead{xin.zhou@yale.edu}

\author[1]{Fan Li}
\cormark[1]
\ead{fan.f.li@yale.edu}



\author%
[1]
{Donna Spiegelman}
\ead{donna.spiegelman@yale.edu}


\cortext[cor1]{Corresponding author}


\begin{abstract}
\textit{Background and objective:}
The stepped wedge cluster randomized trial is a study design increasingly used for public health intervention evaluations. 
Most previous literature focuses on power calculations for this particular type of cluster randomized trials 
for continuous outcomes, along with an approximation to this approach for binary outcomes. Although not accurate for binary outcomes, it has been widely used. To improve the approximation for binary outcomes, two new methods for stepped wedge designs (SWDs) of binary outcomes have recently been published. However, these new methods have not been implemented in publicly available software. The objective of this paper is to present power calculation software for SWDs in various settings for both continuous and binary outcomes.
\\
\textit{Methods:}
We have developed a SAS macro \textbf{\%swdpwr} and an R package \textbf{swdpwr} for power calculation in SWDs. Different scenarios including cross-sectional and cohort designs, binary and continuous outcomes, marginal and conditional models, three link functions, with and without time effects are accommodated in this software.
\\
\textit{Results:}
\textbf{swdpwr} provides an efficient tool to support investigators in the design and analysis of stepped wedge cluster randomized trails. \textbf{swdpwr} addresses the implementation gap between newly proposed methodology and their application to obtain more accurate power calculations in SWDs.
\\
\textit{Conclusions:} 
This user-friendly software makes the new methods more accessible and incorporates as many variations as currently available, which were not supported in other related packages. \textbf{swdpwr} is implemented under two platforms: SAS and R, satisfying the needs of investigators from various backgrounds.
\end{abstract}


\begin{highlights}
\item User-friendly SAS macro and R package for stepped wedge designs power calculations
\item Accommodation of various scenarios for binary and continuous outcomes
\item Addressing the implementation gap and providing more accurate power calculations for binary outcomes
\end{highlights}


\maketitle

\doublespacing
\section{Introduction}\label{sec:intro}

In cluster randomized trials (CRTs), the unit of randomization is the cluster, improving administrative convenience and reducing treatment contamination \citep{murray1998design}. Traditional clustered designs such as the parallel design and the crossover design may be susceptible to ethical concerns, because not all clusters receive the intervention before the end of the study \citep{turner2017review}. In contrast, in stepped wedge designs (SWDs), all clusters start out in the control condition and switch to the intervention condition in a unidirectional and randomly assigned order, and once treated, the clusters maintain their intervention status until the end of the study. At pre-specified time periods, a random subset of clusters cross over from the control to the intervention condition. Stepped wedge randomization may be preferred for estimating intervention effects when it is logistically more convenient to roll-out intervention in a staggered fashion and when the stakeholders or participating clusters perceive the intervention to be beneficial to the target population \citep{hussey2007design}.

Two different stepped wedge designs have been considered: the cross-sectional design and the closed cohort design \citep{copas2015designing}. In a cross-sectional design, different participants are recruited at each time period in each cluster; while in a closed cohort design (which for simplicity will be referred to as a cohort design hereafter), participants are recruited at the beginning of the study and have repeated outcome measures at different time periods \citep{hemming2013efficiency}. A distinguishing feature of all CRTs is that outcomes within the same cluster tend to be correlated with one another \citep{martin2016intra}. Because in SWDs outcomes are measured at different time periods, the within-period and between-period correlation coefficients may be different and thus should be separately considered in designing SWDs \citep{martin2016intra}. An additional within‐individual correlation should be included when it is a cohort SWD to account for the repeated measures within the same individual over time \citep{hughes2015current}. Two statistical models can be used to account for these three levels of intraclass correlation: the conditional model and the marginal model. Conditional models are based on mixed effects models \citep{pinheiro2006mixed, breslow1993approximate}, which accommodate the intraclass correlations via latent random effects. Marginal models describe the population-averaged responses across cluster-periods, and are usually fitted with the generalized estimating equations (GEE) \citep{liang1986longitudinal}. The interpretations of regression parameters can be different under these two models, with the important exception of the identity and log links when random effects and the covariates are independent, as is typically assumed \citep{ritz2004equivalence}. The design and analysis of SWDs have been mostly based on conditional models, for instance, \cite{hussey2007design}; \cite{woertman2013stepped}; \cite{hemming2015stepped}; \cite{hooper2016sample}; \cite{li2018optimal}. As marginal models carry a straightforward population‐averaged interpretation assuming equivalence of studying random effects with target population random effects, \cite{li2018sample} proposed methods for the design and analysis of SWDs using marginal models. Alternatively, the conditional model considers the causal effect of interventions on individual, under the assumption of no unmeasured confounders present or other source of bias.

Binary outcomes are frequently seen in cluster randomized trials as endpoints. However, existing methods for sample size calculation of SWDs have been almost exclusively focused on continuous outcomes. \cite{hussey2007design} proposed an approach based on linear mixed effects models, estimated by weighted least squares for continuous outcomes, and provided an approximation to this approach for binary outcomes. Systematic reviews indicate that the majority of SWDs with binary outcomes used this approximation method \citep{hemming2016sample, martin2016systematic}, which may either overestimate or underestimate the power in different scenarios \citep{zhou2020maximum}. To improve this approximation, \cite{zhou2020maximum} developed a maximum likelihood method for power calculations of SWDs with binary outcomes based on the mixed effects model and \cite{li2018sample} proposed a method for binary outcomes within the framework of GEE that employed a block exchangeable within‐cluster correlation structure with three correlation parameters.

These new methods have not yet been implemented in publicly available software, making it difficult for researchers and practitioners to apply these new methods to design their studies rigorously. Additionally, existing software for SWDs focus on limited settings and do not have accurate power calculations for binary outcomes. \cite{hemming2014menu} developed a Stata menu-driven program \textbf{steppedwedge} based on \cite{hussey2007design}, the \textbf{swCRTdesign} \citep{swCRTdesign} package in R and an Excel spreadsheet (\url{http://faculty.washington.edu/jphughes/pubs.html}) also implemented \cite{hussey2007design}, all of which utilize the linear mixed effects model for continuous outcomes and binary proportions considering only the traditional intra-cluster correlation coefficient for cross-sectional designs assuming fixed time effects. \cite{hemming2020tutorial} developed the \textbf{Shiny} CRT Calculator programmed in R (\url{https://github.com/karlahemming/Cluster-RCT-Sample-Size-Calculator}) using linear mixed effects models with three intraclass correlations that accommodated cross-sectional and cohort designs \citep{hooper2016sample} for continuous outcomes and included the linear mixed effects model approximation for binary outcomes. All these software did not implement accurate power calculations for SWDs with binary outcomes. There is another R package \textbf{SWSamp} (\url{https://sites.google.com/a/statistica.it/gianluca/swsamp}) based on \cite{baio2015sample}, which allowed simulation-based sample size and power calculations for several general scenarios including cross-sectional and cohort designs for continuous, binary and count outcomes. However, this package considered a less complex model and regarded time effects as fixed only. Hence, in an effort to make the new methods more accessible and incorporate as many  variations as currently available, we have developed user-friendly and computationally efficient software based on the methods proposed by \cite{zhou2020maximum} and \cite{li2018sample} to implement power calculations of SWDs for binary and continuous outcomes. The software engine has been developed in Fortran and is incorporated into the SAS macro \textbf{\%swdpwr} and the R package \textbf{swdpwr}.

\section{Methods} \label{sec:models}
Throughout this article, the regression parameter $\beta$ denotes the treatment effect. For testing the treatment effect, we consider the following hypothesis:
\begin{equation} \label{eq:test}
\textrm{H}_0: \beta =  0 \qquad \textrm{versus} \qquad \textrm{H}_A: \beta \neq 0 ,
\end{equation}
where $\beta_A$ is the true value of $\beta$ under the alternative hypothesis that $ \beta_A \neq 0$.
In this software, power is calculated based on a two-sided Wald-type test given by:
\begin{equation} \label{eq:waldpower}
\Phi\left(\frac{|\beta_A|}{\sqrt{\textrm{Var}(\hat{\beta})}}-Z_{1-\alpha/2}\right)+\Phi\left(-\frac{|\beta_A|}{\sqrt{\textrm{Var}(\hat{\beta})}}-Z_{1-\alpha/2}\right),
\end{equation}
where $\Phi(\cdot)$ is the cumulative distribution function of the standard normal distribution, $\alpha$ is the significance level, and $Z_{1-\alpha/2}$ is the $(1-\alpha/2)$th quantile of the standard normal distribution. The variance of $\hat{\beta}$ is defined by either asymptotic theory for maximum likelihood estimation (MLE) in a conditional model, or by the theory of generalized estimating equations in a marginal model. 

A SWD is defined by $I$ clusters and $J$ time periods, each including $K_{ij}$ individuals at time period $j$ for cluster $i$ ($i$ in $1,\cdots,I$; $j$ in $1,\cdots,J$). In a cross-sectional SWD, the size of cluster $i$ is given by $N_i=\sum_{j=1}^JK_{ij}$; in a cohort SWD, assuming that $K_{ij}=K_{i}$ over the active study period, the size of cluster $i$ is $N_i=K_{i}$. Let $Y_{ijk}$ be the response corresponding to individual $k$ at time period $j$ from cluster $i$ ($k$ in $1,\cdots,K_{ij}$), which can be continuous or binary outcomes (for example: success or failure of surgery, getting the disease or not). For both outcome types, we first present the general models with three correlation parameters and then give more details about the cases implemented in this software. More remarks on these methods can be found in Appendix \ref{sec:models3}.

\subsection{Models for binary outcomes}  \label{sec:binary}
Due to the correlated nature of the outcomes in SWDs, a generalized linear mixed effects model can be written:
\begin{equation} \label{eq:timeeffectbinarygeneral}
g(p_{ijk}) = \mu + X_{ij}\beta + \gamma_j +  b_i + c_{ij} + \pi_{ik}
\end{equation}
where $g(\cdot)$ is a link function, $X_{ij}$ is a binary treatment assignment (1=intervention; 0=standard of care) in cluster $i$ at time period $j$, $\mu$ is the baseline outcome rate on the scale of the link function in the control group, 
$\gamma_j$ is the fixed time effect corresponding to time period $j$ ($j$ in $1,\cdots,J$, and $\gamma_1=0$ for identifiability), and $\beta$ is the parameter of interest in this study, the treatment effect. We assume that $b_i$ is the between-cluster random effect distributed by 
$N(0, \sigma_{b}^2)$, $c_{ij}$ is the cluster-by-time interaction random effect distributed by $N(0, \sigma_{c}^2)$, and $\pi_{ik}$ is the random effect for repeated measures of one individual distributed by  $N(0, \sigma_{\pi}^2)$. We assume that $b_i$, $c_{ij}$ and $\pi_{ik}$ are independent of each other.
Let $p_{ijk}=E(Y_{ijk}|X_{ij}, b_i, c_{ij}, \pi_{ik})=Pr(Y_{ijk}=1|X_{ij}, b_i, c_{ij}, \pi_{ik})$ be the probability of the outcome for individual $k$ conditioned on the random effects and design allocation, interpreted as the conditional mean response of the individual.

To account for the correlation of outcomes, a correlation structure with three levels of correlation is employed: (1) $\alpha_{0}$, the within-period correlation, which measures the similarity between responses from different individuals within the same cluster during the same time period (corr$(Y_{ijk},Y_{ijk'})=\alpha_0$ for $k \neq k'$); (2) $\alpha_1$, the between-period correlation, which measures the similarity between responses from different individuals within the same cluster in different time periods (corr$(Y_{ijk},Y_{ij'k'})=\alpha_1$ for $j \neq j', k \neq k'$); (3) $\alpha_2$, the within-individual correlation, which measures the similarity between responses from the same individual across time periods (corr$(Y_{ijk},Y_{ij'k})=\alpha_2$ for $j\neq j'$).

This framework can accommodate three common link functions $g(\cdot)$: identity, log and logit links. In a cross-sectional design, the correlation structure depends only on $\alpha_0$ and $\alpha_1$ since different sets of individuals are included at each time period, and $\alpha_2$ is no longer relevant. However, in this circumstance, $\alpha_1$ and $\alpha_2$ are, by definition, the same, since actually different individuals are assessed at each time period, to obtain a two-level exchangeable correlation model to accommodate the cross-sectional design    \citep{li2018sample}. Otherwise, a cohort design is implied.

 Model (\ref{eq:timeeffectbinarygeneral}) accommodates both cross-sectional and cohort SWDs. These designs can also be fitted by the marginal model with the same correlation structure but without random effects:
 \begin{equation} \label{eq:timeeffectgee}
g(p_{ijk}) = \mu + X_{ij}\beta + \gamma_j.
\end{equation}
  
  In our software, we implemented specific cases of designs with binary outcomes under conditional and marginal model assumptions.
For binary outcomes, the marginal and conditional mean responses limit the ranges of the correlation coefficients inside $[-1,1]$ \citep{qaqish2003family, ridout1999estimating}. Thus, we need additional restrictions for $\alpha_0, \alpha_1, \alpha_2$ beyond those required to ensure a positive definite working correlation matrix and that the correlations are between 0 and 1.  To ensure valid probabilities between 0 and 1 under the identity and log link functions, we also need restrictions for parameters related to the treatment effect, baseline outcome rate on the scale of the link function and time effect parameters. For example, $\mu+\beta+\gamma_j$ should not exceed 0 under log link function. 
These restrictions apply to both the conditional and marginal models. When the input parameters for power calculations are out of range, the software will return an error message. More details will be discussed in Appendix \ref{sec:warnings}.

\subsubsection{Conditional model based on GLMM}
 \cite{zhou2020maximum} considered a cross-sectional SWD modelled by the generalized linear mixed model (GLMM) as a special case of Model (\ref{eq:timeeffectbinarygeneral}):
\begin{equation} \label{eq:timeeffect}
g(p_{ijk}) = \mu + X_{ij}\beta + \gamma_j + b_i.
\end{equation}
All notations are as defined previously. We assume a normal distribution for random effects, $b_i\sim N(0, \tau^2)$, although \cite{zhou2020maximum} showed that results are insensitive to the assumed form of the distribution of the random effects. This model only considers fixed time effects and does not include cluster-by-time interaction random effects. As $\alpha_2$ is undefined for cross-sectional studies, the correlation structure for this case is $ \alpha_0 = \alpha_1$ and \cite{zhou2020maximum} showed that they are equal to $\rho=\frac{Var(E(Y_{ij}|X_{ij}=0,b_i))}{Var(Y_{ij}|X_{ij}=0)}$, where $\rho$ represents an intra-cluster correlation coefficient which measures the correlation between individuals in the same cluster. Under the identity link, $\alpha_0 = \alpha_1  = \frac{\tau^2}{\tau^2+\mu(1-\mu)}$ \citep{hussey2007design}.

 In this work, we consider three link functions for $g(\cdot)$: identity, log and logit. Then, the likelihood for the outcomes is based on the conditional probability under the specific link function. 
Because the likelihood involves integrating out the unobserved random effects $b_i$, we use the Gaussian quadrature for numerical integration \citep{mcculloch1997maximum}. The large-sample variance and power are calculated based on the theory of maximum likelihood estimation.

We also consider SWDs with no time effects (all $\gamma_j=0$). The model is
\begin{equation} \label{eq:notimeeffect}
g(p_{ijk}) = \mu + X_{ij}\beta + b_i.
\end{equation}
It applies when time effects are expected to be minimum \citep{zhou2017cross}. The derivation of the likelihood formula and the calculation of the variance of $\hat{\beta}$ and power are similar to the procedures with time effects.

\subsubsection{Marginal model based on GEE}\label{sec:models2} Cohort SWDs with three correlation parameters, where individuals from each cluster are enrolled at the start of the trial and followed up for repeated measurements \citep{li2018sample}, have been studied. The marginal model based on GEE is:
\begin{equation} \label{eq:timeeffectgee}
g(p_{ijk}) = \mu + X_{ij}\beta + \gamma_j.
\end{equation}
Here we slightly abuse the notation, and denote $p_{ijk}=E(Y_{ijk}|X_{ij})=P(Y_{ijk}=1|X_{ij})$, interpreted as the marginal mean response of individuals, $\mu$ is the marginal outcome rate on the scale of the link function in the control group, $\gamma_j$ is the fixed time effect, $\beta$ is the parameter of interest in this study, interpreted as the marginal treatment effect.   We also consider two settings: with time effects ($j$ in $1,\cdots,J$, and $\gamma_1=0$ for identifiability) and without time effects (all $\gamma_j=0$). When assuming $K_{ij}=K_{i}$ for each $j$, the block exchangeable working correlation matrix for cluster $i$ is written as \citep{li2018sample}:
\begin{equation}  \label{eq:R}
 \textrm R_{i}= (1-\alpha_0 +\alpha_1 - \alpha_2) \textrm I_{JK_{i}} + (\alpha_2 - \alpha_1) \textrm J_{J} \otimes \textrm I_{K_i} + (\alpha_0 - \alpha_1) \textrm I_{J} \otimes \textrm J_{K_i} + \alpha_1 J_{JK_{i}}.
\end{equation}
where $J_u$ is a $u*u$ matrix with all elements of 1, $I_k$ is a $k*k$ identity matrix. \cite{li2018sample} showed that $R_i$ has four distinct eigenvalues and all of them should have positive values, in order to ensure a positive-definite 
correlation structure.
We define $\eta = (\mu, \gamma_2, \gamma_3, ..., \gamma_{J}, \beta)’$ to be the vector of parameters in Model (\ref{eq:timeeffectgee}), and $\eta = (\mu,\beta)’$ to be the parameter vector without time effects. Let $Y_i=(Y_{i11}, Y_{i12}...,Y_{iJK_i})’$ and $p_i=(p_{i11}, p_{i12}...,p_{iJK_i})’$. The GEE estimator $\hat{\eta}$ is solved from $\sum_i \textrm D_i' {\textrm V_i}^{-1} (Y_{i}-p_{i})=0$, where $\textrm D_i = \partial p_i /\partial \eta'$, $\textrm V_i = {A_i}^{1/2} R_i {A_i}^{1/2}$, and ${A_i}$ is the $JK_i-$dimensional diagonal matrix with elements $\phi v(p_{ijk})$, with $\phi$ representing the dispersion parameter ($\phi=1$ for binary outcomes) and the variance function $v(p_{ijk}) = p_{ijk}(1-p_{ijk})$. Assuming the correlation structure is correctly specified, $\hat{\eta}$ is approximately multivariate normal with mean ${\eta}$ and covariance estimated by the model-based  estimator $({\sum_i \textrm D_i'(\hat{\eta}) {\textrm V_i^{-1}({\alpha}}) \textrm D_i(\hat{\eta})})^{-1}$. Hence, we can calculate the variance of $\hat{\beta}$.
 
Cross-sectional SWDs can also be designed under the marginal model with appropriate specification of the correlation parameters $\alpha_0$ and $\alpha_1$.

\subsection{Models for continuous outcomes}\label{sec:conti}
 \cite{hooper2016sample}  assumed a linear mixed effects model for correlated continuous outcomes in SWDs:
\begin{equation} \label{eq:timeeffectcongeneral}
Y_{ijk}= \mu + X_{ij}\beta + \gamma_j +  b_i + c_{ij} + \pi_{ik}+ \epsilon_{ijk}
\end{equation}
Notations are as defined previously. Here we assume that $\epsilon_{ijk}\sim N(0, \sigma_{e}^2)$, and as previously, $b_i$, $c_{ij}$, $\pi_{ik}$ and $\epsilon_{ijk}$ are independent of each other, and the total variance of $Y_{ijk}$ is  $\sigma_{t}^2=\sigma_{b}^2+\sigma_{c}^2+\sigma_{\pi}^2+\sigma_{e}^2$. We note that, corr$(Y_{ijk},Y_{ijk'})=\alpha_0=\frac{(\sigma^2_{b}+\sigma^2_{c})}{\sigma^2_{t}}$ for $k \neq k'$, corr$(Y_{ijk},Y_{ij'k'})=\alpha_1=\frac{\sigma^2_{b}}{\sigma^2_{t}}$ for $j \neq j', k \neq k'$, and corr$(Y_{ijk},Y_{ij'k})=\alpha_2=\frac{(\sigma^2_{b}+\sigma^2_{\pi})}{\sigma^2_{t}}$ for $j\neq j'$. The correlation structure $R_i$ and the specification for cross-sectional and cohort designs are as in Section \ref{sec:binary}.


 This general Model (\ref{eq:timeeffectcongeneral}) agrees with the population-averaged marginal model used by \cite{li2018sample} for SWDs:
\begin{equation} \label{eq:timeeffectlig}
E(Y_{ijk})=\mu_{ijk}=  \mu + X_{ij}\beta + \gamma_j 
\end{equation}
where $\mu_{ijk}$ is the marginal expected mean response of $Y_{ijk}$. Our software allows for SWDs both with time effects ($j$ in $1,\cdots,J$, and $\gamma_1=0$ for identifiability) and without time effects (all $\gamma_j=0$). 

\cite{li2018sample} and \cite{hooper2016sample} both considered scenarios for continuous outcomes that accommodate three correlation parameters, under marginal and conditional model, respectively. In both models, the covariance is estimated by the model-based estimator $(\sum_{i}\bf{Z'_{i}}\bf{V_{i}^{-1}}\bf{Z_{i}})^{-1}$, where $\bf{Z_{i}}$ is the $JK_i*(J+1)$ design matrix corresponds to the parameter vector $\eta$ under the SWD framework within cluster $i$ and $\bf{V_i}$ is the working covariance matrix within cluster $i$ based on a block exchangeable correlation structure, in which $\bf{V_i}$ $=\sigma_{t}^2{R_i}$. Our software implements \cite{li2018sample} for  cross-sectional and cohort SWDs with continuous outcomes. Regaring restrictions on the allowable parameter space for continuous outcomes, 
the correlations are within $[-1,1]$. When two or three correlation parameters are used for cross-sectional and cohort designs, respectively, a positive definite correlation matrix $R_i$ should be ensured.

     \begin{table}
              
      \begin{tabular}{c|c|c c}
      \hline
   \multicolumn{2}{c}  {}  & Conditional model &Marginal model\\\hline
\multirow{6}{*}{ \tabincell{c}{Binary outcomes\\(identity, log, logit links)}}     & \multirow{3}{*}{Cross-sectional design }       & \multirow{3}{*}{\cite{zhou2020maximum}}
        & \multirow{3}{*}{\cite{li2018sample}}  \\
        & \multirow{3}{*}    &\multirow{3}{*} & \\
         & \multirow{3}{*}    &\multirow{3}{*} &  \\\cline{2-4}
                                                          & \multirow{3}{*} {Cohort design }      & \multirow{3}{*}{N/A}
         &\multirow{3}{*}{\cite{li2018sample}}   \\
           & \multirow{3}{*}    &\multirow{3}{*} &  \\
           & \multirow{3}{*}    &\multirow{3}{*} & \\\hline
       \multirow{6}{*}{\tabincell{c}{Continuous outcomes\\(identity link)}}     & \multirow{3}{*}{Cross-sectional design } & \multicolumn{2}{c}{\multirow{3}*{\tabincell{c}{\cite{hussey2007design,li2018sample,zhou2017cross}  }}}\\
        & \multirow{3}{*}    &\multirow{3}{*}  &\\
         & \multirow{3}{*}    &\multirow{3}{*} &\\\cline{2-4}
                                                            &\multirow{3}{*}{Cohort design }    &  \multicolumn{2}{c}{\multirow{3}*{\cite{hooper2016sample,li2018optimal,li2018sample}}} \\
        & \multirow{3}{*}    &\multirow{3}{*} &   \\
        & \multirow{3}{*}    &\multirow{3}{*} &\\\hline
 \end{tabular}
     \caption{Methods implemented in {\bf{swdpwr}}.}
             \label{listmethod}
      \end{table}

\section{Software Description} \label{sec:illustrations}
Table \ref{listmethod} displays all the scenarios that are implemented in the software, accommodating cases and methods with and without time effects.
The input parameters are the same for R and SAS. Different input parameters values, including for the anticipated mean response rate in the control group at the start of the study, the anticipated mean response rate in the control group at the end of the study, the anticipated mean response rate in the intervention group at the end of the study, the study design, and the intraclass correlations can be specified based on the preliminary data.  The arguments of software \textbf{swdpwr} are shown in Table \ref{tab2}.

 In this section, we give a detailed description for the implementation in R and SAS. Examples under different scenarios will be discussed in Section \ref{sec:examples}. We assume balanced cluster-period sizes so that $K_{ij}=K$ over all $i$ and $j$. 
\subsection{Specification of regression model parameters}\label{sec:illustrations1}
To account for time effects, we assume that the difference of time effects between neighbouring time periods is constant, which implies that the absolute value of $\gamma_j$ increases linearly in $j$. Hence, parameter for time effects $\gamma_J$ is a single parameter. This software considers two complementary ways 
for users to provide the regression model parameters $\mu$, $\beta$ and $\gamma_J$.
One approach is to specify \verb+meanresponse_start+, \verb+meanresponse_end0+ and \verb+meanresponse_end1+, which will help identify the baseline outcome rate on the scale of the link function $\mu$, the treatment effect $\beta$ and the time effect $\gamma_J$  on the link function scale. The alternative approach is to specify \verb+meanresponse_start+, \verb+meanresponse_end0+ and \verb+effectsize_beta+. This approach provides the treatment effect $\beta$ directly, while $\mu$ and $\gamma_J$ can also be identified from the input parameters. 

It is discussed in \cite{li2018sample} that the power calculations for continuous outcomes do not depend on $\mu$ and $\gamma_J$, hence users are allowed to get the results by just providing the accurate parameter \verb+effectsize_beta+ if it is continuous outcome and assumes presence of time effects. However, users will additionally need to specify different values for \verb+meanresponse_start+ and \verb+meanresponse_end0+ to indicate the presence of time effects, although the values do not require to be accurate. If assuming no time effects and continuous outcomes, Equation (\ref{eq:timeeffectvar2no}) in Appendix \ref{appendixA2} also shows the independence of power on $\mu$, and users can only supply \verb+effectsize_beta+ to conduct the calculation, with the default of no time effects.

In case of any contradiction of parameter specification, either one of the approaches for parameter specification is allowed in the software only. Users are not allowed to supply \verb+meanresponse_start+, \verb+meanresponse_end0+, \verb+meanresponse_end1+ and \verb+effectsize_beta+ at the same time.

\subsection{Description of the R package}
This package \textbf{swdpwr}  provides statistical power calculation for SWDs with both binary and continuous outcomes through the \verb+swdpower+ function. The arguments of \verb+swdpower+ are
\begin{lstlisting}[language=R, caption=]
swdpower(K, design, family = "binomial", model = "conditional", link = "identity", 
type = "cross-sectional", meanresponse_start = NA, meanresponse_end0 = meanresponse_start, 
meanresponse_end1 =  NA, effectsize_beta = NA, sigma2 = 0, typeIerror = 0.05, alpha0 = 0.1, 
alpha1 = alpha0/2, alpha2 = NA)

\end{lstlisting}
with details provided in Table \ref{tab2} along with defaults if any (those above with an equal sign have default values). The input argument\verb+ design+ is a matrix generated in R with elements of 0 (control) or 1 (intervention) for users to define the detail of the SWD, in which each row represents a cluster and each column represents a time period. The argument \verb+sigma2+ is only allowed for continuous outcomes. The argument \verb+alpha2+ should not be an input under cross-sectional designs although it is numerically identical to \verb+alpha1+ in this scenario by definition. The object returned by \verb+swdpower+ function has a class of \verb+swdpower+, which includes a list of the design matrix, summary features of the design and the power for the scenario specified. This package also contains an associated print method for the class of \verb+swdpower+, which hides the list output and just shows the readable power under the alternative hypothesis.

\subsection{Description of the SAS macro}\label{sec:illustrations3}
The SAS macro \textbf{\%swdpwr} plays the same role as the function \verb+swdpower+ in R package \textbf{swdpwr}  described above, and this macro also needs a pre-prepared data set for the study design matrix generated in SAS. For illustration, a toy data set \verb+design+  is created, which contains the allocation of control (0) or intervention (1) for each cluster at different time periods as well as the number of clusters for each allocation shown the first column. The step to generate this data set in SAS is as follows, in which each type of allocation is conducted in 3 clusters.
\begin{lstlisting}[language=SAS, caption=]
data design;
input numofclusters time1 time2 time3 time4;
cards;
3 0 1 1 1
3 0 0 1 1
;
run;
\end{lstlisting}
The procedure of data set generation is the common \verb+DATA+ step, but is supposed to include a header with correct time period description.

Sharing the same utility with the function  \verb+swdpower+  in R package \textbf{swdpwr}, the input arguments of \textbf{\%swdpwr} are
\begin{lstlisting}[language=SAS, caption=]
% macro swdpwr(K, design, family = "binomial", model = "conditional", link = "identity", 
type = "cross-sectional", meanresponse_start = NA, meanresponse_end0 = meanresponse_start, 
meanresponse_end1 =  NA, effectsize_beta = NA, sigma2 = 0, typeIerror = 0.05, alpha0 = 0.1, 
alpha1 = alpha0/2, alpha2 = NA)
\end{lstlisting}
with details given in Table \ref{tab2} together with defaults if any. This macro will return an output of summary features of the design, as well as the power value under the particular scenario, which can be referred in Appendix \ref{sec:applications}.

\begin{table}[]
       \centering
\begin{tabular}{lll}

     \hline
 \multicolumn{1}{l}{Argument} & \multicolumn{1}{l}{Description} & \multicolumn{1}{l}{ Default }    \\\hline
\verb+K+   & \tabincell{l}{number of individuals at each time period\\ in a cluster}     &  \\ \hline
\verb+design+   &  \tabincell{l}{I*J dimensional data set that describes \\the study design}     &    \\  \hline
\verb+family+   &    \tabincell{l}{specify family="gaussian" for continuous \\outcome, family="binomial" for binary \\outcome}        &\verb+"binomial"+  \\ \hline
\verb+model+    &     \tabincell{l}{specify model="conditional" for \\conditional model, model="marginal" for \\marginal model}                  &\verb+"conditional"+ \\ \hline
\verb+link+  &     \tabincell{l}{choose link function from link="identity", \\link="log" and link="logit"}           &\verb+"identity"+  \\ \hline
\verb+type+   &     \tabincell{l}{specify type="cohort" for closed cohort \\ study, 
type="cross-sectional"  for \\ cross-sectional  study}           &\verb+"cross-sectional"+ \\ \hline
\verb+meanresponse_start+    & \tabincell{l}{ the anticipated mean response rate in the \\control group at the start of the study} &  \verb+NA+\\ \hline
\verb+meanresponse_end0+   &  \tabincell{l}{the anticipated mean response rate in the \\ control group at the end of the study}     & \verb+meanresponse_start+  \\ \hline
\verb+meanresponse_end1+    &  \tabincell{l}{the anticipated mean response rate in the \\ intervention group at the end of the study}     & \verb+NA+\\ \hline
\verb+effectsize_beta+    &  \tabincell{l} { the anticipated effect size}         & \verb+NA+ \\ \hline
\verb+sigma2+    &  \tabincell{l} {marginal variance of the outcome\\ (only allowed if  is continuous outcome)}         & \verb+0+ \\ \hline
\verb+typeIerror+   &  Type I error            & \verb+0.05+\\ \hline
\verb+alpha0+    & within-period correlation   $\alpha_0$         &\verb+0.1+ \\ \hline
\verb+alpha1+  &  between-period  correlation   $\alpha_1$      &  \verb+alpha0/2+\\ \hline
\verb+alpha2+   &  \tabincell{l}{within-individual correlation $\alpha_2$\\ (only allowed if type="cohort")}          & \verb+NA+ \\
     \hline
\end{tabular}
 \caption{Arguments of software \textbf{swdpwr}.}
              \label{tab2}
\end{table}

\section{Examples} \label{sec:examples}
The usage of the software is based on platforms of R and SAS, which requires separate illustrations. The following sections 
are organized according to different scenarios such as continuous and binary outcomes, cross-sectional and cohort settings, different model options, different link functions, different time effects assumptions. Each section will contain examples under both platforms. When the input arguments are inappropriate, warnings and error messages may occur in the software. Appendix \ref{sec:warnings} will give an illustration of these scenarios and advise users how to correct for the errors. Besides the hypothetical examples in this section, we provide two real world applications implemented by SAS in Appendix \ref{sec:applications}: the Washington Expedited Partner Therapy (EPT) trial and the Tanzania postpartum intrauterine device (PPIUD) study. 

\subsection{Conditional model with binary outcome and cross-sectional design}
\cite{zhou2020maximum} proposed this maximum likelihood method under cross-sectional settings, which discussed scenarios with time effects and without time effects. With binary outcomes, the variance argument \verb+sigma2+ is undefined and should not be specified. Due to this specific cross-sectional setting, we specify $\alpha_0$ = $\alpha_1$ in the function and $\alpha_2$ is also undefined as discussed in Section \ref{sec:models2}.
\subsubsection{R package}
The study design should be supplied by an I*J matrix, where I is the number of clusters, J is the number of time periods. We will utilize the \verb+rep()+ function to specify the elements of the study design matrix by providing  different allocations and repeated times of each allocation. More details can be found in the following examples.
Fitting the model with all required inputs in R:
\begin{lstlisting}[language=R, caption=]
R> library("swdpwr")
R> dataset = matrix(c(rep(c(0, 1, 1), 6), rep(c(0, 0, 1), 6)), 12, 3, byrow = TRUE)
R> swdpower(K = 50, design = dataset, family = "binomial",  model = "conditional", 
+ link = "identity", type = "cross-sectional", meanresponse_start = 0.2, 
+ meanresponse_end0 = 0.25, meanresponse_end1 = 0.38, typeIerror = 0.05, alpha0 = 0.01, 
+ alpha1 = 0.01)
This cross-sectional study has total sample size of 1800
Power for this scenario is 0.899 for the alternative hypothesis treatment effect
beta = 0.13 ( Type I error =  0.05 )     
\end{lstlisting}

In the example above, 50 individuals are included in each cluster at each time period, with 12 clusters and 3 time periods under the SWD for a total sample size of 1800. The calculation was conducted utilizing the conditional model with time effects and the identity link function. The intraclass correlations, here interpreted as the correlation between different individuals within the same cluster, were all set as 0.01 and the Type I error was 0.05. The power obtained from \verb+swdpower+ for this scenario is 0.899 for the alternative hypothesis $\beta_A=0.13$.

The model can also be considered under other link functions such as log and logit link functions. For instance:
\begin{lstlisting}[language=R, caption=]
R> library("swdpwr")
R> dataset = matrix(c(rep(c(0, 1, 1), 6),rep(c(0, 0, 1), 6)), 12, 3, byrow = TRUE)
R> swdpower(K = 50, design = dataset, family = "binomial", model = "conditional", 
+ link = "logit", type = "cross-sectional", meanresponse_start = 0.2,  
+ meanresponse_end0 = 0.25, meanresponse_end1 = 0.38, typeIerror = 0.05, alpha0 = 0.01,  
+ alpha1 = 0.01)
This cross-sectional study has total sample size of 1800
Power for this scenario is 0.838 for the alternative hypothesis treatment effect 
beta = 0.616 ( Type I error =  0.05 )          
\end{lstlisting}
This example assumes a conditional model with time effects and the logit link function, with other inputs the same as the above example.  R package gives the power of 0.838 for the alternative hypothesis $\beta_A=0.616$.

\subsubsection{SAS macro}
 The maro \textbf{\%swdpwr}  is used in SAS. The same examples are illustrated in SAS as above. As in R, \textbf{\%swdpwr} in SAS requires that the input of a design matrix and a SAS data set \verb+exmple+ as follows is a pre-requisite:
\begin{lstlisting}[language=SAS, caption=]
data example;
input numofclusters time1 time2 time3;
cards;
6 0 1 1
6 0 0 1
;
run;
\end{lstlisting}

The call to \textbf{\%swdpwr} is:
\begin{lstlisting}[language=SAS, caption=]
%swdpwr(K = 50, design = example, family = "binomial", model = "conditional", 
link = "identity", type = "cross-sectional", meanresponse_start = 0.2, 
meanresponse_end0 = 0.25,  meanresponse_end1 = 0.38, typeIerror = 0.05,
alpha0 = 0.01, alpha1 = 0.01)
\end{lstlisting}


\begin{lstlisting}[language=SAS, caption=]
%swdpwr(K = 50, design = example, family = "binomial", model = "conditional", 
link = "logit", type = "cross-sectional", meanresponse_start = 0.2, 
meanresponse_end0 = 0.25,  meanresponse_end1 = 0.38, typeIerror = 0.05, 
alpha0 = 0.01, alpha1 = 0.01)
\end{lstlisting}


The data set \verb+example+ should be generated as described in Section \ref{sec:illustrations3}. The inputs of the SAS are the same as the R examples.
Calculations were based on the conditional model with time effects. The same results are obtained as in the R examples above. SAS gives the power of 0.899 under this scenario with the identity link function for the alternative hypothesis $\beta_A=0.13$, and the power of 0.838 under this scenario with the logit link function for the alternative hypothesis $\beta_A=0.616$.

If considering scenarios without time effects, users can provide the same value for
\verb+meanresponse_start+ and \verb+meanresponse_end0+.

\subsection{Marginal model for binary outcomes with cross-sectional and cohort designs} \label{sec:margianlbinary}
A method for the marginal model based on GEE was proposed in \cite{li2018sample}, employing a block exchangeable correlation structure applicable to both cohort and cross-sectional designs. This method accommodates scenarios with and without time effects as well as three link functions. When the outcome is binary, because the variance is a strict function of the mean, the marginal variance argument \verb+sigma2+ is undefined.

\subsubsection{R package with cohort design}
We first utilize a cohort design example to illustrate the use of the software in R for function \verb+swdpower+ in package \textbf{swdpwr}. A cohort design requires the specification of $\alpha_0$, $\alpha_1$ and $\alpha_2$.
\begin{lstlisting}[language=R, caption=]
R> library("swdpwr")
R> dataset = matrix(c(rep(c(0, 1, 1, 1), 6), rep(c(0, 0, 1, 1), 6)), 12, 4, byrow = TRUE)
R> swdpower(K = 100, design = dataset, family = "binomial", model = "marginal",
+ link = "log", type = "cohort", meanresponse_start = 0.156, meanresponse_end0 = 0.1765,  
+ effectsize_beta = 0.75, typeIerror = 0.05, alpha0 = 0.03, alpha1 = 0.015, alpha2 = 0.2)
This cohort study has total sample size of 1200
Power for this scenario is 0.983 for the alternative hypothesis treatment effect
beta = 0.75 ( Type I error =  0.05 )        
\end{lstlisting}

This example illustrates a cluster randomized cohort SWD including 100 individuals in each cluster, with 12 clusters and 4 time periods for a total sample size of 1200. We conducted the calculation using the marginal model under the log link function with time effects. The three correlation parameters were 0.03, 0.015 and 0.2. The power calculated given by R for this scenario is around 0.983 for the alternative hypothesis $\beta_A=0.75$.

Different link functions can be considered for the calculation. For instance, with the same regression model parameters and other inputs but under the logit link function, the power obtained is then 0.843 for the alternative hypothesis $\beta_A=0.75$.
\begin{lstlisting}[language=R, caption=]
R> library("swdpwr")
R> dataset = matrix(c(rep(c(0, 1, 1, 1),  6), rep(c(0, 0, 1, 1), 6)), 12, 4, byrow = TRUE)
R> swdpower(K = 100, design = dataset, family = "binomial",  model = "marginal",
+ link = "logit", type = "cohort", meanresponse_start = 0.1349, meanresponse_end0 = 0.1499,  
+ effectsize_beta = 0.75, typeIerror = 0.05, alpha0 = 0.03, alpha1 = 0.015, alpha2 = 0.2)
This cohort study has total sample size of 1200
Power for this scenario is 0.843 for the alternative hypothesis treatment effect 
beta = 0.75 ( Type I error =  0.05 )    
\end{lstlisting}

\subsubsection{SAS macro with cohort design}
Next, we illustrate the use of SAS macro \textbf{\%swdpwr} for cohort design examples and a SAS data set \verb+design2+ is required:
\begin{lstlisting}[language=SAS, caption=]
data design2;
input numofclusters time1 time2 time3 time4;
cards;
6 0 1 1 1
6 0 0 1 1
;
run;
\end{lstlisting}
To implement the same examples in R as above, the macro called to do power calculations would be:
\begin{lstlisting}[language=SAS, caption=]
% swdpwr(K = 100, design = design2, family = "binomial", model = "marginal", 
link = "log", type = "cohort", meanresponse_start = 0.156,
meanresponse_end0 = 0.1765, effectsize_beta = 0.75, typeIerror = 0.05,
alpha0 = 0.03, alpha1 = 0.015, alpha2 = 0.2)
\end{lstlisting}
\begin{lstlisting}[language=SAS, caption=]
% swdpwr(K = 100, design = design2, family = "binomial", model = "marginal", 
link = "logit", type = "cohort", meanresponse_start = 0.1349,
meanresponse_end0 = 0.1499, effectsize_beta = 0.75, typeIerror = 0.05, 
alpha0 = 0.03, alpha1 = 0.015, alpha2 = 0.2)
\end{lstlisting}

Based on the sample stepped wedge cohort trial and conducting design using the marginal model with time effects, 
we obtain the same calculated power as above: power of 0.983 under the log link function and power of 0.843 under the logit link function for the alternative hypothesis $\beta_A=0.75$.

\subsubsection{R package with cross-sectional design}
We next consider the cross-sectional design under the marginal model, where $\alpha_0$ and $\alpha_1$ should be specified. The example conducted by \verb+swdpower+ in R package \textbf{swdpwr}:
\begin{lstlisting}[language=R, caption=]
R> library("swdpwr")
R> dataset = matrix(c(rep(c(0, 1, 1, 1), 6), rep(c(0, 0, 1, 1), 6)), 12, 4, byrow = TRUE)
R> swdpower(K = 100, design = dataset, family = "binomial", model = "marginal", 
+ link = "identity", type = "cross-sectional", meanresponse_start = 0.15, 
+ meanresponse_end0 = 0.15, meanresponse_end1  = 0.2, typeIerror = 0.05, alpha0 = 0.02, 
+ alpha1 = 0.015)
This cross-sectional study has total sample size of 4800
Power for this scenario is 0.946 for the alternative hypothesis treatment effect 
beta = 0.05 ( Type I error =  0.05 )            
\end{lstlisting}
Above is a cross-sectional SWD with 100 individuals at each time period in each cluster for a total sample size of 4800. The design was conducted on the marginal model under identity link function without time effects. In this example, the powe given by R is around 0.946 for the alternative hypothesis $\beta_A=0.05$.

\subsubsection{SAS macro with cross-sectional design}
To illustrate the cross-sectional settings in SAS, we use the same example in R as above. Similarly, we use the data set \verb+design2+ generated previously as the study design.

The macro called to do power calculations would be:
\begin{lstlisting}[language=R, caption=]
%swdpwr(K = 100, design = design2, family = "binomial", model = "marginal", 
link = "identity", type = "cross-sectional", meanresponse_start = 0.15,
meanresponse_end0 = 0.15,  meanresponse_end1  = 0.2, typeIerror = 0.05,
alpha0 = 0.02, alpha1 = 0.015)
\end{lstlisting}

With a total sample size of 4800, the power under this scenario is 0.946 for the alternative hypothesis $\beta_A=0.05$, which accords with the results in R.



\subsection{Method with continuous outcomes under identity link}
Linear mixed effects models have been widely used in the design and analysis of SWDs, for example, see: \cite{hussey2007design}, \cite{hemming2015stepped}, \cite{hooper2016sample}. \cite{li2018sample} proposed methods for analyzing SWDs for continuous outcomes under the identity link function. Because it has been shown that conditional and marginal models are equivalent when the regression is linear with mean 0 random effect and the identity or log link functions \citep{ritz2004equivalence}, with the addition of a new proof for equivalence of variance of treatment effect in Appendix \ref{appendixA}, the procedures in \cite{li2018sample} are used for power calculations with continuous outcomes, with three levels of correlation parameters are considered.

The designs for continuous outcomes are conducted under the identity link function. Both cross-sectional and cohort designs with and without time effects can be accommodated in our software. Here, the argument \verb+sigma2+ must be specified, which equals $\sigma_t^2$ as in Section \ref{sec:conti}. 

\subsubsection{R package}
An example from \cite{li2018sample} is used for illustrating the use of function \verb+swdpower+ in R:
\begin{lstlisting}[language=R, caption=]
R> library("swdpwr")
R> dataset = matrix(c(rep(c(0, 1, 1),4), rep(c(0, 0, 1), 4)), 8, 3, byrow = TRUE)
R> swdpower(K = 24, design = dataset, family = "gaussian", model = "marginal", 
+ link = "identity", type = "cohort", meanresponse_start = 0.1, meanresponse_end0 = 0.2, 
+ effectsize_beta = 0.2, sigma2 = 0.095,  typeIerror = 0.05, alpha0 = 0.03, alpha1 = 0.015, 
+ alpha2 = 0.2) 
This cohort study has total sample size of 192 
Power for this scenario is 0.965 for the alternative hypothesis treatment effect 
beta = 0.2 ( Type I error =  0.05 ) 
\end{lstlisting}

In this example trial, the outcome is continuous with total marginal variance of 0.095, conducted in 8 clusters and 3 time periods for a total sample size of 192. Marginal model under identity link function with time effects is utilized for the calculation. This trial is in a cohort design with three different levels of correlation parameter. The power obtained from the software is around 0.965 for the alternative hypothesis $\beta_A=0.2$.

We also consider the same example without time effects:

\begin{lstlisting}[language=R, caption=]
R> library("swdpwr")
R> dataset = matrix(c(rep(c(0, 1, 1), 4), rep(c(0, 0, 1), 4)), 8, 3, byrow = TRUE)
R> swdpower(K = 24, design = dataset,  family= "gaussian", model = "marginal",  
+ link = "identity", type = "cohort", effectsize_beta = 0.2, sigma2 = 0.095,  
+ typeIerror = 0.05, alpha0 = 0.03, alpha1 = 0.015, alpha2 = 0.2)
This cohort study has total sample size of 192 
Power for this scenario is 1 for the alternative hypothesis treatment effect 
beta = 0.2 ( Type I error =  0.05 )  
\end{lstlisting}

In this example, only \verb+effectsize_beta+ is specified and the trial was conducted without time effects by default. The calculated power is around 1 for the alternative hypothesis $\beta_A=0.2$.

\subsubsection{SAS macro}
Next, we illustrate the R examples above in macro \textbf{\%swdpwr} using SAS, and prepare a SAS data set \verb+design3+: 
\begin{lstlisting}[language=SAS, caption=]
data design3;
input numofclusters time1 time2 time3;
cards;
4 0 1 1
4 0 0 1
;
run;
\end{lstlisting}

The macro called to do power calculations would be:
\begin{lstlisting}[language=SAS, caption=]
%swdpwr(K = 24, design = design3, family = "gaussian", model = "marginal",
link = "identity", type = "cohort", meanresponse_start = 0.1,
meanresponse_end0 = 0.2, effectsize_beta = 0.2, sigma2 = 0.095,
typeIerror = 0.05, alpha0 = 0.03, alpha1 = 0.015, alpha2 = 0.2)

%swdpwr(K = 24, design = design3, family = "gaussian", model = "marginal",
link = "identity", type = "cohort", effectsize_beta = 0.2, sigma2 = 0.095, 
typeIerror = 0.05, alpha0 = 0.03, alpha1 = 0.015, alpha2 = 0.2)
\end{lstlisting}

The same results were obtained as in R, with the power of around 0.965 considering time effects and 1 considering no time effects, respectively, for the alternative hypothesis $\beta_A=0.2$.

\section{Discussion} \label{sec:summary}
This article has described the use of the R package \textbf{swdpwr} and the  SASmacro \textbf{\%swdpwr} for power calculations in SWDs. The software is designed under two computer platforms where users specify input parameters for different scenarios of interest, accommodating cross-sectional and cohort designs, binary and continuous outcomes, marginal and conditional models, three link functions, with and without time effects. This software addresses the implementation gap between newly proposed methodology and their application to obtain more accurate power calculation for binary outcomes in SWDs, rather than relying on the approximation method in \cite{hussey2007design}. The core of the software is developed in Fortran to ensure efficient computations, and is linked to SAS and R by the foreign function interface. As the number of time periods and individuals in each cluster increases, the calculations may become time-consuming, thus we have implemented some methods to decrease the computational cost. For example, we applied a simplified closed-form expression for matrix inversion under the marginal model, as derived in \cite{li2018sample}. The SAS macro \textbf{swdpwr} can be accessed at: \url{https://github.com/jiachenchen322/swdpwr}. Package \textbf{swdpwr} is available via the Comprehensive R Archive Network (CRAN) as a contributed package at: \url{https://CRAN.R-project.org/package=swdpwr}. The package is also available as a \textbf{shiny} \citep{shiny}  app, available online at: \url{https://jiachenchen322.shinyapps.io/swdpwr_shinyapp/} to enable users without SAS or R programming skills to also use this software easily.

In the future, we plan to accommodate various numbers of individuals per cluster and even at each time period to make the software even more flexible. Furthermore, we are working to implement the 
partition method developed in \cite{zhou2020maximum} for the conditional model with binary outcomes to permit power and sample size calculations with cluster sizes, $K$, greater than 150 as occurs in practice. In addition, current software only accommodates nested exchangeable and block exchangeable correlation structures, which could be further extended to allow for damped exponential correlation structures \citep{munoz1992parametric} and proportional decay correlation structures \citep{li2020design}.

\section*{Declaration of Competing Interest}
The authors have declared no conflict of interest.

\section*{Acknowledgments}
This work was supported by the grants NIH/R01AI112339 and NIH/DP1ES025459.

\bibliographystyle{elsarticle-num}
\bibliography{cas-refs}

\singlespacing
  \renewcommand{\theequation}{\thesection.\arabic{equation}}
  \setcounter{equation}{0}  
\renewcommand\thefigure{\thesection.\arabic{figure}}    
\setcounter{figure}{0}    
\renewcommand\thetable{\thesection.\arabic{table}}    
\setcounter{table}{0}    
\appendix
\section*{Appendix}
\section{Remarks for methods}\label{sec:models3}
This software includes procedures for power calculations in SWDs with binary outcomes under the identity, logit and log link functions, based on \cite{zhou2020maximum} and \cite{li2018sample}, and also incorporates procedures for continuous outcomes as in \cite{hussey2007design}, \cite{hooper2016sample} and \cite{li2018sample} with the identity link function. The treatment effect $\beta$ is the same under the conditional and marginal model with the identity and log link functions \citep{ritz2004equivalence}.

In Appendix \ref{appendixA}, by proving the equivalence of the variance of the intervention effect in the conditional and marginal models with continuous outcomes, we showed that \cite{li2018sample} incorporated the cross-sectional designs in \cite{hussey2007design} and the cohort designs in \cite{hooper2016sample}, and obtained the same power formulae for the conditional and marginal models for each of these designs. Hence, we can directly implement \cite{li2018sample} in our software for all of these continuous outcome scenarios.

\begin{figure}

\centering 

\begin{minipage}{7cm}
\subfigure[Identity link, without time effects.]{\label{fig:subfig:a}

\includegraphics[width=1\linewidth]{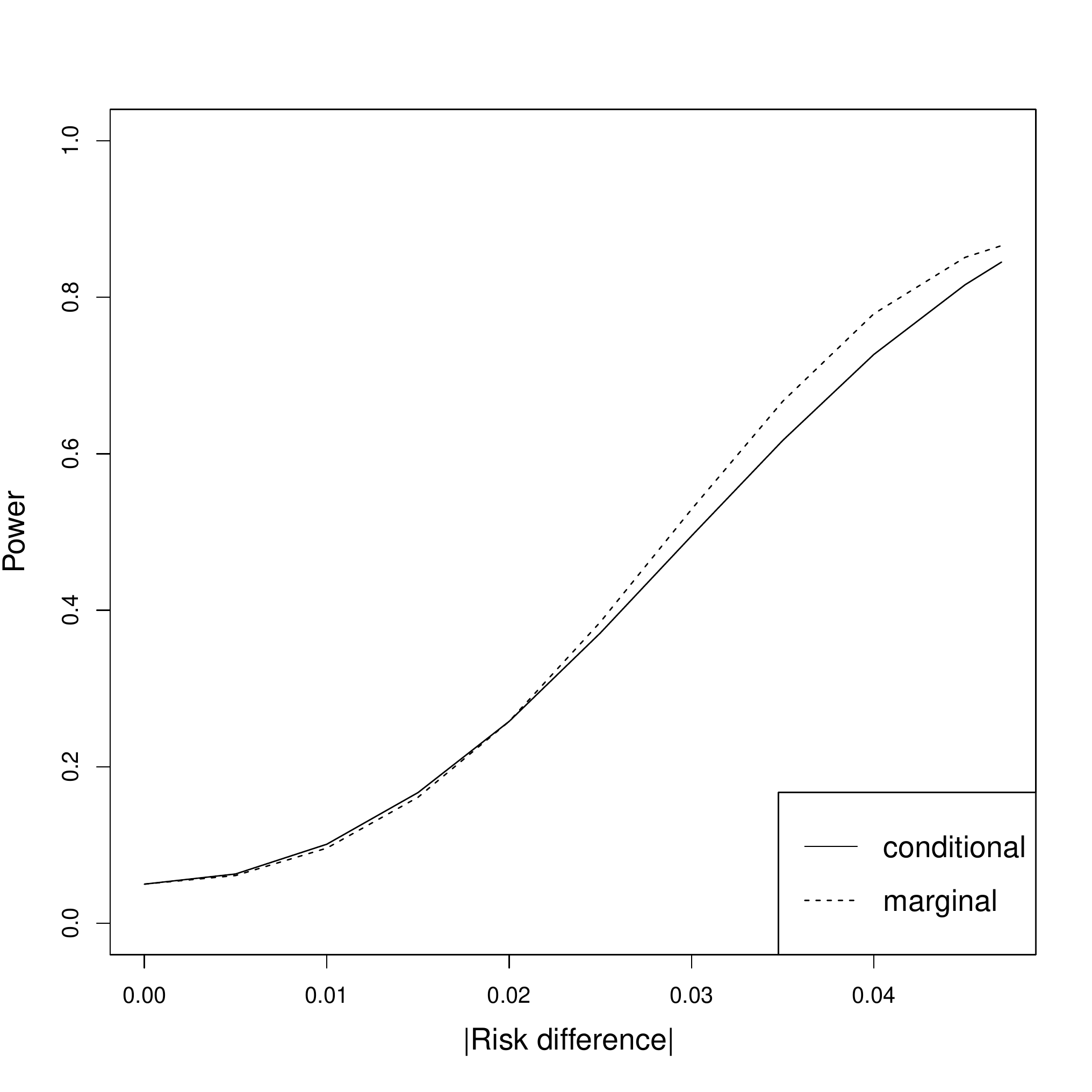}}
\end{minipage}
\begin{minipage}{7cm}
\subfigure[Log link, without time effects.]{\label{fig:subfig:b}

\includegraphics[width=1\linewidth]{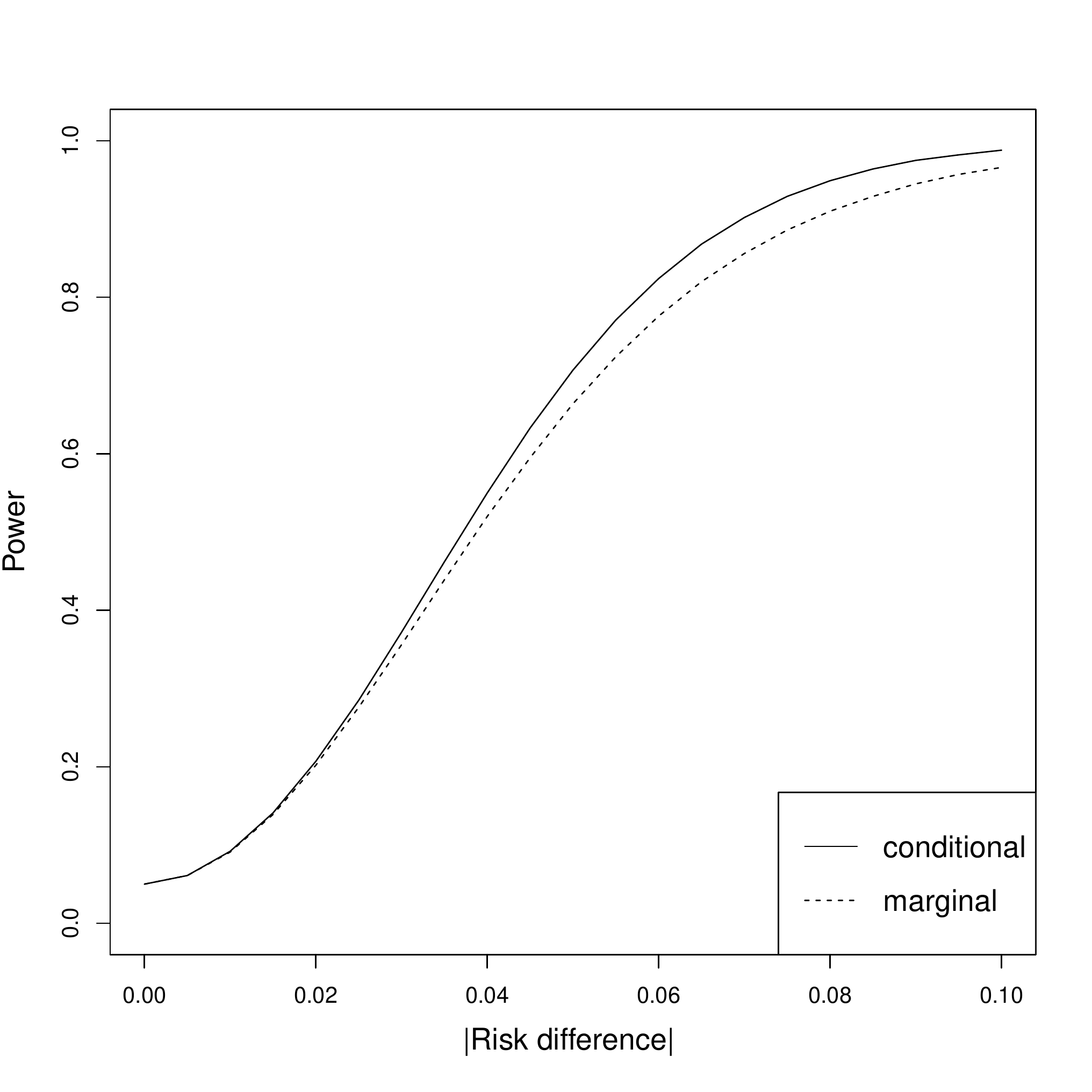}}
\end{minipage}

\begin{minipage}{7cm}
\subfigure[Logit link, with time effects.]{\label{fig:subfig:c}

\includegraphics[width=1\linewidth]{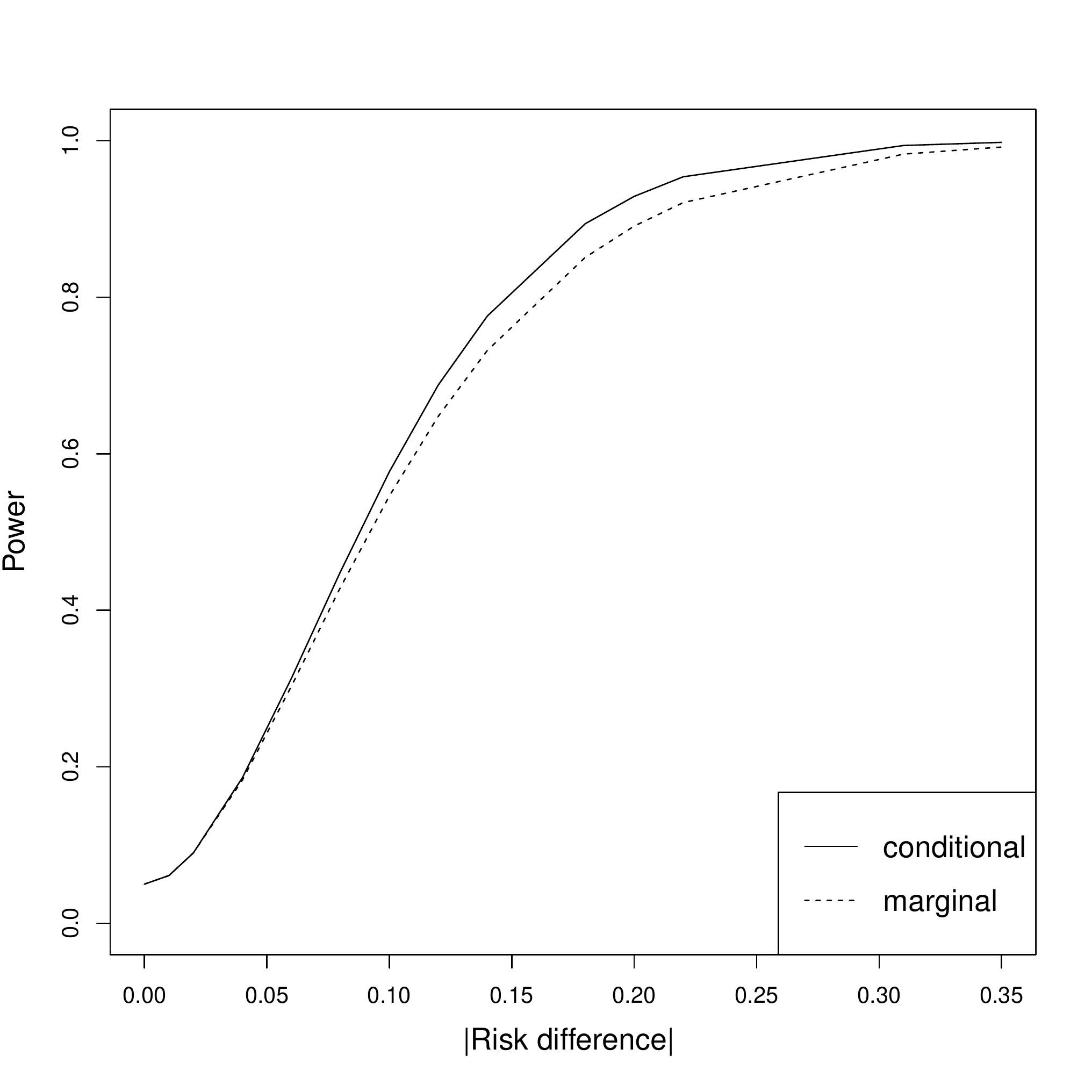}}
\end{minipage}
\begin{minipage}{7cm}
\subfigure[Logit link, with time effects.]{\label{fig:subfig:d}

\includegraphics[width=1\linewidth]{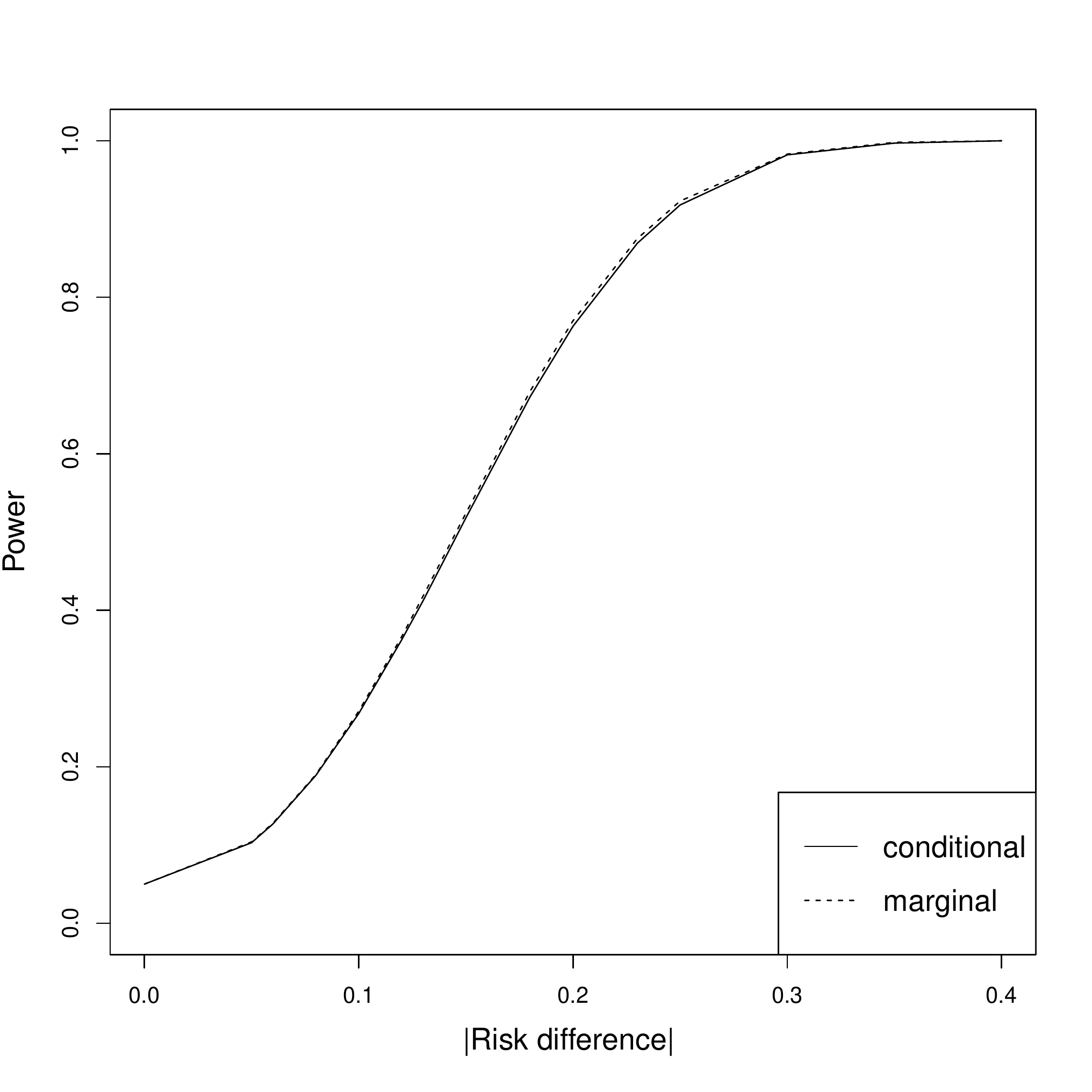}}
\end{minipage}
\caption{Power vs. the absolute value of risk difference between the control and intervention group at the end of the cross-sectional study, for I = 6, J = 3, K = 40, and $\alpha_0=\alpha_1=0.02$.  There are no time effects in the first row. Figure \ref{fig:subfig:a} has the identity link and the baseline mean response rate in the control group of 0.05. The conditional model has a slightly larger power when the  absolute value of risk difference is smaller, the opposite conclusion is shown when the absolute value of risk difference is larger. Figure \ref{fig:subfig:b} has the log link and the baseline mean response rate in the control group of 0.045. The conditional and marginal models have almost the same power when risk difference is smaller, and conditional model has a larger power when risk difference increases.  The second row considers time effects and the logit link. In Figure \ref{fig:subfig:c}, the baseline mean response rate in the control group is 0.03 and the mean response rate in the control group at the end of the study is 0.04. The conditional model has a larger power as risk difference increases. In Figure \ref{fig:subfig:d}, the baseline mean response rate in the control group is 0.3 and the mean response rate in the control group at the end of the study is 0.4. The marginal model has a larger power as risk difference increases.
}
\label{fig:subfig}
\end{figure}

Regarding the cross-sectional settings where $\alpha_0 = \alpha_1 $ with binary outcomes, we implement both the conditional \citep{zhou2020maximum} and the marginal \citep{li2018sample} models, for which the variance of  these two estimates are different. As shown by \cite{ritz2004equivalence}, the treatment effect, $\beta$, is the same for these two models under the identity and log link functions given the same mean response rates. However, the estimates of the conditional and marginal model have different  variances, and we found in numerical calculations that neither is globally greater than the other. Under the logit link, the parameter $\beta$ has different interpretations for the two models given the same mean response rates, and again, neither has a greater power than the other although variance comparisons are not meaningful with parameters having different interpretations. Figure \ref{fig:subfig} provides some examples of this pattern, which is expected because GLMM is likelihood based, while the marginal models relies on unbiased estimating equations.

We emphasize that our software considers designs with and without time effects. In most literature, see, for instance, \cite{hussey2007design}, \cite{hemming2015stepped}, \cite{hooper2016sample}, and \cite{li2018optimal,li2018sample}, time effects are assumed. However, \cite{zhou2017cross} argued that SWDs are mostly used in the study of relatively short-term outcomes with short-term interventions, thus, it is reasonable to assume no time effects in the primary analysis. \cite{zhou2020maximum} developed their method considering cases with and without time effects. When developing this software, we include and generalize all the methods to cases with and without time effects. 

\section{Warnings and error messages generation}\label{sec:warnings}
The program will output reliable results although warnings are returned due to some invalid input arguments, including mainly five scenarios. First, conditional model with binary outcomes only allows for cross-sectional settings, so if \verb+"cohort"+ is specified for the argument  \verb+type+, a warning message will occur and \verb+type=“cross-sectional”+ will be forced to resume the power calculation. Second, $\alpha_0 =\alpha_1$ is to be ensured for conditional model with binary outcome and cross-sectional designs, if it is violated, the value of $\alpha_1$  is set to the value of $\alpha_0$ and a warning will be returned. Third, for cross-sectional designs, if $\alpha_2$ is specified, it will be ignored and a warning message will remind users that  $\alpha_2$ is undefined and should not be an input. Fourth, when  \verb+sigma2+ is supplied for binary outcomes, it will be ignored and the software will return a warning that marginal variance should not be specified for binary outcomes. Last, if it is continuous outcome supplied by link functions of logit or log, a warning message will explain that the program will be conducted with the link function forced to be identity.

There are also cases where error messages occur and the program will be stopped, then users will need to revise the input arguments. The most common errors will be missing of input arguments. However, users will need to notice that although the software includes two complementary ways to supply regression model parameters as in Section \ref{sec:illustrations1}, an error will be generated if users specify  \verb+meanresponse_start+, \verb+meanresponse_end0+, \verb+meanresponse_end1+ and \verb+effectsize_beta_+ simultaneously, in case that contradictions occur for the value of parameters. Besides, the specification of input arguments cannot be identified when typos occur in arguments such as \verb+family+, \verb+model+, \verb+link+ and \verb+type+, then errors will be reported.

When input parameters, including the correlation parameters, Type I error rate, or response for binary outcomes, are out of range, an error message will be returned. Before giving the detailed examples, we discuss  the specification of correlation parameters for both cross-sectional and cohort designs. As described in Section \ref{sec:models2}, these correlations represent the within-period, between-period, and within-individual correlations, respectively, with the natural restriction of within $[-1,1]$. For the correlation structure of SWDs, 
\cite{li2018optimal} identified four linear eigenvalue constraints to ensure a positive definite correlation matrix. These constraints are enforced in our software for all models and both outcome types. In a cross-sectional design, because there are no repeated measures within subjects, $\alpha_2$ is not required and the block exchangeable correlation structure reduces to the nested exchangeable structure as in \cite{teerenstra2010sample}. Thus, in cross-sectional designs, a single intra-cluster correlation coefficient which measures the correlation between individuals in the same cluster describes the correlation structure in \cite{zhou2020maximum} such that the working correlation reduces to  $\alpha_0=\alpha_1$. 

Additional restrictions for $\alpha_0$, $\alpha_1$ and $\alpha_2$ are needed as the marginal and conditional means limit the ranges of correlation due to binary outcomes \citep{qaqish2003family}, where $i \neq j$:
\begin{equation} \label{eq:restrict1}
max(0, \mu_{i}+\mu_{j}-1) \leq E[Y_{i} Y_{j}] \leq  min(\mu_{i},\mu_{j})  
\end{equation}
This can be written to require that, the following three criteria must be satisfied over $i = 1,2, ..., I; j, l = 1,2,...,J; k, m = 1,2,...,K$:

\begin{equation} \label{eq:restrict2}
E[Y_{ijk} Y_{ijm}] = \mu_{ij} \mu_{ij} +\alpha_0 v_{ij}^{1/2}v_{ij}^{1/2}
\end{equation}
\begin{equation} \label{eq:restrict3}
E[Y_{ijk} Y_{ilm}] = \mu_{ij} \mu_{il} +\alpha_1 v_{ij}^{1/2}v_{il}^{1/2}
\end{equation}
\begin{equation} \label{eq:restrict4}
E[Y_{ijk} Y_{ilk}] = \mu_{ij} \mu_{il} +\alpha_2 v_{ij}^{1/2}v_{il}^{1/2}
\end{equation}
where $\mu_{ijk}=\mu_{ij}$ and $v_{ijk}=v_{ij}$ are the mean and variance of the outcome for individual $k$ at time period $j$ from cluster $i$, and $j \neq l$ plus $k \neq m$. These restrictions apply to both the conditional and marginal model with binary outcomes.

Now, we give examples for scenarios where representative errors occur in R, so that users can check the input parameter according to the error message and obtain suggestions about how to revise them. The same errors will occur if conducted in SAS as well.

In the first example, the requirement that the 2- or 3-way working correlation matrix $R_i$ is positive definite is violated as correlation parameters exceed plausible ranges although they are still within $[0,1]$. This error could occur for  both outcome types and in all models.
\begin{lstlisting}[language=R, caption=]
R> library("swdpwr")
R> dataset = matrix(c(rep(c(0, 1, 1, 1), 4), rep(c(0, 0, 1, 1), 4), 
+ rep(c(0, 0, 0, 1), 4)), 12, 4, byrow = TRUE)
R> swdpower(K = 100, design = dataset, family = "gaussian", 
+ model = "marginal", link = "identity", type = "cohort",
+ effectsize_beta = 0.05, sigma2 = 0.095, typeIerror = 0.05, alpha0 = 0.015, 
+ alpha1 = 0.2, alpha2 = 0.1)
Error in swdpower(K = 100, design = dataset, family = "gaussian",
model = "marginal", : Correlation matrix R is not positive definite. Please
check whether the between-period correlation is unrealistically larger 
than the within-period correlation or the within-individual correlation.
\end{lstlisting}

The fix to this error can be revising \verb+alpha1+ to be much smaller, then the program will be processed without errors:
\begin{lstlisting}[language=R, caption=]
R> swdpower(K = 100, design = dataset, family = "gaussian", 
+ model = "marginal", link = "identity", type = "cohort", 
+ effectsize_beta = 0.05, sigma2 = 0.095, typeIerror = 0.05, alpha0 = 0.015, 
+ alpha1 = 0.01, alpha2 = 0.1)
This cohort study has total sample size of 1200
Power for this scenario is 0.994 for the alternative hypothesis treatment
effect beta = 0.05 ( Type I error =  0.05 )  
\end{lstlisting}

The second example occurred with a binary outcome. This error is also related to the correlations values given because they are additionally restricted by  means due to binary outcomes \citep{qaqish2003family, ridout1999estimating} as discussed before.

\begin{lstlisting}[language=R, caption=]
R> library("swdpwr")
R> dataset = matrix(c(rep(c(0, 1, 1, 1), 4), rep(c(0, 0, 1, 1), 4), 
+ rep(c(0, 0, 0, 1), 4)), 12, 4, byrow = TRUE)
R> swdpower(K = 100, design = dataset, family = "binomial", 
+ model = "marginal", link = "identity", type = "cohort", 
+ meanresponse_start = 0.1, meanresponse_end0 = 0.2, effectsize_beta = 0.7, 
+ typeIerror = 0.05, alpha0 = 0.1, alpha1 = 0.05, alpha2 = 0.2)
Error in swdpower(K = 100, design = dataset, family = "binomial", 
model = "marginal", : Correlation parameters do not satisfy the 
restrictions of Qaqish (2003). Please check whether it is possible to reduce
the effect size, or make adjustments to the intraclass correlations.
\end{lstlisting}

This error can be fixed by revising the intraclass correlations, for example:
\begin{lstlisting}[language=R, caption=]
R> swdpower(K = 100, design = dataset, family = "binomial",
+ model = "marginal", link = "identity", type = "cohort", 
+ meanresponse_start = 0.1, meanresponse_end0 = 0.2, effectsize_beta = 0.7, 
+ typeIerror = 0.05, alpha0 = 0.05, alpha1 = 0.05, alpha2 = 0.1) 
This cohort study has total sample size of 1200
Power for this scenario is 1 for the alternative hypothesis treatment
effect beta = 0.7 ( Type I error =  0.05 )  
\end{lstlisting}

The third example also pertains to a binary outcome. Sometimes, the input parameters related to mean responses may exceed the range for valid probability between $[0, 1]$. This error is easy to handle with and is not relevant to continuous outcomes. For instance:
\begin{lstlisting}[language=R, caption=]
R> library("swdpwr")
R> dataset = matrix(c(rep(c(0, 1, 1, 1),  4), rep(c(0, 0, 1, 1), 4), 
+ rep(c(0, 0, 0, 1), 4)), 12, 4, byrow = TRUE)
R> swdpower(K = 100, design = dataset, family = "binomial", 
+ model = "marginal", link = "identity", type = "cohort",
+ meanresponse_start = 0.1, meanresponse_end0 = 0.2, effectsize_beta = 0.9, 
+ typeIerror = 0.05, alpha0 = 0.05, alpha1 = 0.05, alpha2 = 0.1)
Error in swdpower(K = 100, design = dataset, family = "binomial", 
model = "marginal", : Violation of valid probability, given input parameters: 
max(meanresponse_start, meanresponse_end0, meanresponse_end1)>1. Please
check whether any of these values are out of range and revise one or more of 
them.
\end{lstlisting}

The fourth example occurs only in scenarios of binary outcomes under conditional models with time effects. It is discussed in \cite{zhou2020maximum} that as the sample size $K$ for each cluster at each time period increases, the estimated running time increases prohibitively to beyond what is possible even for high performance computing under these scenarios. Hence, we set an upper limit of 150 for $K$ in the current software. However, even with values of $K$ within this allowable range, computing time may still be low. To address this, users can consider switching to the marginal model specification or dropping time effects. An example for this error is:
\begin{lstlisting}[language=R, caption=]
R> library("swdpwr")
R> dataset = matrix(c(rep(c(0, 1, 1, 1), 4), rep(c(0, 0, 1, 1), 4), 
+ rep(c(0, 0, 0, 1), 4)), 12, 4, byrow = TRUE)
R> swdpower(K = 160, design = dataset, family = "binomial", 
+ model = "conditional", link = "logit", type = "cross-sectional", 
+ meanresponse_start = 0.1, meanresponse_end0 = 0.2, effectsize_beta = 0.6, 
+ typeIerror = 0.05, alpha0 = 0.05, alpha1 = 0.05)
Error in swdpower(K = 100, design = dataset, family = "binomial",
model = "conditional", : K should be at least smaller than 150 for this
scenario as the running time is too long with this K for the power
calculation of binary outcomes under conditional model with time effects.
Please reduce K or use the model without time effects or use
marginal models.
\end{lstlisting}

In addition, the intraclass correlations and Type I error must be between 0 and 1 as well for all models, for instance:
\begin{lstlisting}[language=R, caption=]
R> library("swdpwr")
R> dataset = matrix(c(rep(c(0, 1, 1, 1), 4), rep(c(0, 0, 1, 1), 4),
+ rep(c(0, 0, 0, 1), 4)), 12, 4, byrow = TRUE)
R> swdpower(K = 100, design = dataset, family = "binomial", 
+ model = "marginal", link = "identity", type = "cohort", 
+ meanresponse_start = 0.1, meanresponse_end0 = 0.2, effectsize_beta = 0.5,
+ typeIerror = 0.05, alpha0 = 1.1, alpha1 = 0.05, alpha2 = 0.1)
Error in swdpower(K = 100, design = dataset, family = "binomial", 
model = "marginal",  : Violate range of intraclass correlations: 
max(alpha0,alpha1,alpha2)>1. Please correct the values of correlation 
parameters, they must be between 0 and 1.

R> swdpower(K = 100, design = dataset, family = "binomial", 
+ model = "marginal", link = "identity", type="cohort",
+ meanresponse_start = 0.1, meanresponse_end0 = 0.2, effectsize_beta = 0.5,
+ typeIerror = 1.05, alpha0 = 0.1, alpha1 = 0.05, alpha2 = 0.1)
Error in swdpower(K = 100, design = dataset, family = "binomial", 
model = "marginal",  : Type I error provided is larger than 1, it must be
between 0 and 1, and is usually 0.05.

\end{lstlisting}

\section{Application} \label{sec:applications}
The Washington Expedited Partner Therapy (EPT) trial was a community-based trial employing a cluster randomized SWD for promoting EPT. The outcome was Chlamydia status, a binary variable. 24 local health jurisdictions (LHJs) were included in this trial and each represented a cluster. There were 5 time periods and the intervention was initiated at four of them, with 6 clusters entering the intervention group at each time period. We can describe this design in the SAS data set \verb+ept+:
\begin{lstlisting}[language=SAS, caption=]
data ept;
input numofclusters time1 time2 time3 time4 time5;
cards;
6 0 1 1 1 1
6 0 0 1 1 1
6 0 0 0 1 1
6 0 0 0 0 1
;
run;
\end{lstlisting}
 The design used a generalized linear mixed model under the log link function with covariates for intervention status and time period \citep{golden2015uptake}. This cross-sectional design assumes 162 individuals in each cluster at each time period for a total sample size of 19440. Based on preliminary data, the baseline prevalence of Chlamydia was about 0.05, the cluster effect coefficient of variation was 0.3, the Type I error was set of 0.05, and a prevalence ratio of 0.7 was to be tested. The coefficient of variation is defined to characterize the cluster effects on the variance of responses \citep{hayes1999simple} and is closely related to regular intraclass correlation. 
 Following \cite{hussey2007design} that the coefficient of variation was $\tau/\mu'$, where  $\mu'$ is denoted as the baseline prevalence of 
Chlamydia 0.05, the intraclass correlation $\frac{\tau^2}{\tau^2+\mu'(1-\mu')}$ was approximately 0.0047 and hence in this study $\alpha_0=\alpha_1=0.0047$. We also assumed the presence of time effects. The input parameters were calculated and summarized based on these information. \cite{hussey2007design} used the conditional model and the approximation method for binary outcomes, here, we conducted this design using the marginal model, as follows: 
\begin{lstlisting}[language=SAS, caption=]
%swdpwr(K = 162, design = ept, family = "binomial", model = "marginal", 
link = "log", type = "cross-sectional", meanresponse_start = 0.05,
meanresponse_end0 = 0.049, meanresponse_end1 = 0.035, typeIerror = 0.05, 
alpha0 = 0.0047, alpha1 = 0.0047)
\end{lstlisting}

We obtained power of 0.812 for the alternative hypothesis $\beta_A=-0.336$ from the marginal model in Table \ref{tab7}, which corresponds to the anticipated power of 80\% from similar scenarios with the condiitonal model considered in \cite{golden2015uptake} and \cite{hussey2007design}.

\begin{table}[]
       \centering
\begin{tabular}{lll}

     \hline
 \multicolumn{1}{l}{Result}    \\\hline
 I =  24     \\
 J =  5   \\
 K =  162     \\
 Total sample size = 19440\\
 Family =  binomial  \\
 Model =  marginal  \\
 Link =   log  \\
 Type = cross-sectional \\
 Baseline (mu): -2.996  \\
Treatment effect (beta): -0.336
   \\
Time effect (gamma J): -0.020  \\
 alpha0: 0.005 \\
 alpha1: 0.005   \\
 alpha2: 0.005  \\
 Type I error =  0.050    \\
 Power = 0.812   \\
     \hline
\end{tabular}
\caption{Output from macro \textbf{\%swdpwr} for EPT trial}
              \label{tab7}
\end{table}

The Tanzania PPIUD study utilized a SWD to assess the causal effect of a PPIUD intervention on subsequent pregnancy \citep{canning2016institutionalizing}. The binary outcome was defined as a current or terminated pregnancy at 18 months postpartum. 6 hospitals were selected into the trial and the study lasted for 18 months with 4 time periods. We can describe this design in a SAS data set \verb+PPIUD+:
\begin{lstlisting}[language=SAS, caption=]
data PPIUD;
input numofclusters time1 time2 time3 time4;
cards;
3 0 1 1 1
3 0 0 0 1
;
run;
\end{lstlisting}

A generalized linear mixed model under identity link without time effects was employed for design. For illustrative purposes, we consider a small cluster size and this cross-sectional design assumes 120 individuals in each cluster at each time period for a total sample size of 2880. As in \cite{canning2016institutionalizing}, the baseline proportion of pregnancy was assumed to be 0.24, the intraclass correlation was 0.15, the Type I error was set of  0.05, and a prevalence ratio of around 0.8 was to be tested. We estimated from preliminary data that the effect size was -0.046, assuming no time effects. We conducted this design using the conditional model in SAS. The macro called for this power calculation would be:
\begin{lstlisting}[language=SAS, caption=]
%swdpwr(K = 120, design = PPIUD, family = "binomial", model = "conditional",
link = "identity", type = "cross-sectional", meanresponse_start = 0.24, 
meanresponse_end0 = 0.24, effectsize_beta = -0.046, typeIerror = 0.05, 
alpha0 = 0.15, alpha1 = 0.15)
\end{lstlisting}

\begin{table}[]
       \centering
\begin{tabular}{lll}

     \hline
 \multicolumn{1}{l}{Result}    \\\hline
I =  6     \\
 J =  4   \\
 K =  120    \\
 Total sample size = 2880\\
 Family =  binomial \\
 Model =  conditional   \\
 Link =  identity  \\
 Type = cross-sectional\\
 Baseline (mu):  0.240  \\
 Treatment effect (beta): -0.046
   \\
Time effect (gamma J): 0.000  \\
 alpha0: 0.150   \\
 alpha1: 0.150 \\
 alpha2: 0.150 \\
 Type I error =  0.050   \\
 Power = 0.846      \\
     \hline
\end{tabular}
    \caption{Output from macro \textbf{\%swdpwr} for PPIUD study }
              \label{tab8}
\end{table}

We get the  power 0.846 for the alternative hypothesis $\beta_A=-0.046$ from the conditional model in Table \ref{tab8}. This result is consistent with the conclusion of 80\% power or more in \cite{canning2016institutionalizing}.

\section{Equivalence of the variance for continuous outcomes}\label{appendixA}
In this paper, we include procedures for power calculations in SWDs with continuous outcomes  based on \cite{li2018sample}, which employed a block exchangeable correlation structure with three correlation parameters accommodating both cohort and cross-sectional designs under the GEE framework. \cite{hussey2007design} developed methods based on a linear mixed effects model with a continuous outcome for a cross-sectional design with a single random cluster effect. \cite{hooper2016sample} generalized \cite{hussey2007design}'s method to closed cohort CRTs with random effects of individuals within cluster, between- and  within-period effect, as in \cite{li2018sample}. 

Here, we show the equivalence of
power obtained by \cite{hussey2007design,hooper2016sample,li2018sample} 
for a continuous outcome. Then, power calculation for continuous outcomes under all scenarios covered in this software can all be incorporated by \cite{li2018sample} directly.

\subsection{With time effects} \label{appendixA1}
With $I$ clusters and $J$ time periods, \cite{hussey2007design} defined:

\begin{equation} \label{eq:timeeffecthussey0}
Y_{ijk}=  \mu + X_{ij}\beta + \gamma_j + b_i + \epsilon_{ijk}
\end{equation}
where $Y_{ijk}$ is the individual continuous response  in cluster $i$ at time period $j$ of individual $k$, $X_{ij}$ is a binary treatment assignment (1=intervention; 0=standard of care) in cluster $i$ at time period $j$, $\beta$ is the treatment effect, $\mu$ is the baseline outcome rate on the scale of the link function in control groups, $\gamma_j$ is the fixed time effect corresponding to time period $j$ (with $\gamma_1=0$), $b_i$ is the random effect for cluster $i$ with $b_i\sim N(0, \tau^2)$, and $\epsilon_{ijk}\sim N(0, \sigma_{e}^2)$. We also assume  that $\epsilon_{ijk}$ is independent of $b_i$, and $K$ individuals at each time period in each cluster.

It can be shown that, the cluster means model obtained by summing over individuals within a cluster is:
\begin{equation} \label{eq:timeeffecthussey1}
Y_{ij}=  \mu + X_{ij}\beta + \gamma_j + b_i + \epsilon_{ij}
\end{equation}
where $\epsilon_{ij}= \sum_{k} \epsilon_{ijk}/K \sim N(0, \sigma^2)$ and $\sigma^2=\sigma^2_{e}/K$.

The estimate of the fixed parameter vector $\eta = (\mu, \gamma_2, \gamma_3, ..., \gamma_J, \beta)$ is obtained by weighted least squares (WLS). We assume that $\bf{Z}$ is the $IJ*(J+1)$ design matrix corresponding to the parameter vector $\eta$ under SWD. To obtain power, we are interested in the variance of $\hat{\beta}$, which is the $(J+1) 
\times (J+1)$ element of $(\bf{Z'}\bf{V^{-1}}\bf{Z})^{-1}$, where $\bf{V}$ is an $IJ*IJ$ block diagonal matrix that measures the covariance of mean response between different time periods in all clusters. With K individuals at each time period per cluster,  \cite{hussey2007design} gave:

\begin{equation} \label{eq:timeeffectvar1}
Var(\hat{\beta})= \frac{I\sigma^2(\sigma^2+J\tau^2)}{(IU-W)\sigma^2+(U^2+IJU-JW-IV)\tau^2}
\end{equation}
where $U=\sum_{ij} X_{ij}$, $W=\sum_{j}(\sum_{i}X_{ij})^2$ and $V=\sum_{i}(\sum_{j}X_{ij})^2$.


To generalize cross-sectional SWDs considered by \cite{hussey2007design} to cohort SWDs, \cite{hooper2016sample} also assumed a linear mixed effects model:
\begin{equation} \label{eq:timeeffecthopper}
Y_{ijk}= \mu + X_{ij}\beta + \gamma_j +  b_i + c_{ij} + \pi_{ik}+ \epsilon_{ijk}
\end{equation}
The notations are the same as Section \ref{sec:conti}. We also assume that $b_i$, $c_{ij}$, $\pi_{ik}$ and $\epsilon_{ijk}$ are independent of each other, and the total variance of $Y_{ijk}$ is $\sigma_{t}^2=\sigma_{b}^2+\sigma_{c}^2+\sigma_{\pi}^2+\sigma_{e}^2$. The working correlation matrix is the same as the three-level correlation structure $R_i$ given in Section \ref{sec:conti} for cross-sectional and cohort designs. Model (\ref{eq:timeeffecthopper}) corresponds to individual-level responses based on mixed effects model, which agrees with the population-averaged marginal model in \cite{li2018sample}:
\begin{equation} \label{eq:timeeffectli}
E(Y_{ijk})=\mu_{ijk}=  \mu + X_{ij}\beta + \gamma_j 
\end{equation}
where $\mu_{ijk}$ is the mean response of $Y_{ijk}$ under both the marginal and conditional model. \cite{li2018sample} and \cite{hooper2016sample} are developed to accommodate the same three correlation parameters, under marginal and conditional models, respectively. In both models, the covariance is estimated by the model-based estimator $(\sum_{i}\bf{Z'_{i}}\bf{V_{i}^{-1}}\bf{Z_{i}})^{-1}$, where $\bf{Z_{i}}$ is the $JK_i*(J+1)$ design matrix corresponding to the parameter vector $\eta$ under SWD within cluster $i$ and $\bf{V_i}$ $=\sigma_{t}^2{R_i}$ is the working covariance matrix within cluster $i$ based on a block exchangeable correlation structure. Hence, the estimators of $\beta$ under Model (\ref{eq:timeeffecthopper}) and (\ref{eq:timeeffectli}) are equivalent and obtain the same power.

As shown in \cite{li2018sample}, this variance for the estimated intervention effect is:
\begin{equation} \label{eq:timeeffectvar2}
Var(\hat{\beta})= \frac{(\sigma_t^2/K)IJ\lambda_{3}\lambda_{4}}{(U^2+IJU-JW-IV)\lambda_{4}-(U^2-IV)\lambda_3}
\end{equation}
where  $U=\sum_{ij} X_{ij}$, $W=\sum_{j}(\sum_{i}X_{ij})^2$, $V=\sum_{i}(\sum_{j}X_{ij})^2$, $\lambda_3 = 1+(K-1)(\alpha_0-\alpha_1)-\alpha_2$, $\lambda_4 = 1+(K-1)\alpha_0+(J-1)(K-1)\alpha_1+(J-1)\alpha_2$. 

The equivalence of Model (\ref{eq:timeeffecthussey1}) and (\ref{eq:timeeffectli}) is proved under the cross-sectional setting in \cite{hussey2007design} with $\alpha_0=\alpha_1=\alpha_2=\frac{\tau^2}{\sigma^2_{e}+\tau^2}$, where Equation (\ref{eq:timeeffectvar1}) is equal to Equation (\ref{eq:timeeffectvar2}). 
The treatment effect $\beta$ is the same under the conditional and marginal model with the identity link function \citep{ritz2004equivalence}.
Thus we can conclude the equivalence of these three models in cases with time effects.

\subsection{Without time effects}\label{appendixA2}
Here, we consider cases without fixed time effects and derive the three models by \cite{hussey2007design,hooper2016sample,li2018sample}  accordingly.

The \cite{hussey2007design} model without time effects is denoted as:
\begin{equation} \label{eq:timeeffecthusseynotime}
Y_{ijk}=  \mu + X_{ij}\beta  +  b_i + \epsilon_{ijk}
\end{equation}
The notations are the same as previously.

The cluster means model obtained by summing over individuals within a cluster is:
\begin{equation} \label{eq:timeeffecthussey1notime}
Y_{ij}=\mu + X_{ij}\beta +  b_i + \epsilon_{ij}
\end{equation}
where $\epsilon_{ij}= \sum_{k} \epsilon_{ijk}/K \sim N(0, \sigma^2)$ and $\sigma^2=\sigma^2_{e}/K$.

The estimate of the fixed parameter vector $\eta = (\mu,\beta)$ is obtained by weighted least squares (WLS). We assume that $\bf{Z}$ is the $IJ*2$ design matrix corresponding to the parameter vector $\eta$ under SWD. To obtain power, we are interested in the variance of $\hat{\beta}$, which is the $2\times 2$ element of $(\bf{Z'}\bf{V^{-1}}\bf{Z})^{-1}$, where $\bf{V}$ $=\sigma_{t}^2{R_i}$ is an $IJ*IJ$ block diagonal matrix that measures the covariance of mean response between different time periods in all clusters. With K individuals at each time period per cluster, \cite{liao2015note} showed that the variance of the estimated intervention effect is:

\begin{equation} \label{eq:timeeffectvar1no}
Var(\hat{\beta})= \frac{IJ(\sigma^2+J\tau^2)\sigma^2}{(IJU-U^2)\sigma^2+IJ(JU-V)\tau^2}
\end{equation}
where $U=\sum_{ij} X_{ij}$ and $V=\sum_{i}(\sum_{j}X_{ij})^2$.


The model that accounts for closed cohort designs without time effects under \cite{hooper2016sample} is:
\begin{equation} \label{eq:timeeffecthopperno}
Y_{ijk}= \mu + X_{ij}\beta +  b_i + c_{ij} + \pi_{ik}+ \epsilon_{ijk}
\end{equation}
The notations are defined in the same way as in the previous section. This model is described at individual-level responses based on mixed effects model, which still agrees with the population-averaged marginal model in \cite{li2018sample}:

\begin{equation} \label{eq:timeeffectlino}
E(Y_{ijk})=\mu_{ijk}= \mu + X_{ij}\beta
\end{equation}
where $\mu_{ijk}$ is the mean response of $Y_{ijk}$ under both the marginal and conditional model. \cite{li2018sample} and \cite{hooper2016sample} are developed to accommodate the three three correlation parameters, under marginal and conditional models, respectively. In both models, the covariance is estimated by the model-based estimator $(\sum_{i}\bf{Z'_{i}}\bf{V_{i}^{-1}}\bf{Z_{i}})^{-1}$, where $\bf{Z_{i}}$ is the $JK_i*2$ design matrix corresponding to the parameter vector $\eta$ under SWD  within cluster $i$ and $\bf{V_i}$ is the working covariance matrix within cluster $i$ based on a block exchangeable correlation structure. Hence, Model (\ref{eq:timeeffecthopperno}) and (\ref{eq:timeeffectlino}) are equivalent and obtain the same power.

According to our derivation following \cite{li2018sample}, the variance of the estimated intervention effect under cases without time effects is:
\begin{equation} \label{eq:timeeffectvar2no}
Var(\hat{\beta})= \frac{(\sigma_{t}^2/K)IJ\lambda_{3}\lambda_{4}}{(IJU-IV)\lambda_{4}-(U^2-IV)\lambda_3}
\end{equation}
where  $U=\sum_{ij} X_{ij}$ and $V=\sum_{i}(\sum_{j}X_{ij})^2$, $\lambda_3 = 1+(K-1)(\alpha_0-\alpha_1)-\alpha_2$, $\lambda_4 = 1+(K-1)\alpha_0+(J-1)(K-1)\alpha_1+(J-1)\alpha_2$. 

The equivalence of Model (\ref{eq:timeeffecthussey1notime}) and (\ref{eq:timeeffectlino}) is proved under the cross-sectional setting in \cite{hussey2007design} with $\alpha_0=\alpha_1=\alpha_2=\frac{\tau^2}{\sigma^2_{e}+\tau^2}$, where Equation (\ref{eq:timeeffectvar1no}) equals to Equation (\ref{eq:timeeffectvar2no}). 
 The treatment effect $\beta$ is the same under the conditional and marginal model with the identity link function \citep{ritz2004equivalence}.
Thus we can conclude the equivalence of these three models in cases without time effects.




\end{document}